\documentstyle[epsf,rotate,prl,aps]{revtex}

\begin{document}

\title{Optimal Path in Two and Three Dimensions}

\author{Nehemia~Schwartz, Alexander~L.~Nazaryev, and Shlomo~Havlin}

\address{Minerva Center and Department of Physics,
	 Jack and Pearl Resnick Institute of Advanced Technology Bldg.,
         Bar--Ilan~University,
         52900~Ramat--Gan, Israel}

\maketitle

\bigskip

\begin{abstract}
We apply the Dijkstra algorithm to generate optimal paths
between two given sites on a lattice representing a disordered energy
landscape.  We study the geometrical and energetic scaling properties of
the optimal path where the energies are taken from a uniform distribution. 
Our numerical results for both two and three dimensions suggest that the
optimal path for random uniformly distributed energies is in the same
universality class as the directed polymers.
We present physical realizations of polymers in disordered energy landscape for which
this result is relevant.  

\end{abstract}

\narrowtext

Recently, there have been much interest in the problem of finding
the optimal path in a disordered energy landscape. The optimal path can
be defined as follows.
Consider a $d$-dimensional lattice, where each bond is assigned with
a random energy value taken from a given distribution. The optimal path between
two sites is defined as the path on which the sum of the energies
is minimal. This problem is of relevance to various fields, such as
spin glasses
\cite{Mezard/Parisi/Sourlas/Toulouse/Virasoro:1984},
protein folding
\cite{Ansari/Berendzen/Bowne/Fraunfelder/Iben/Sauke/Shyamsunder/Young:1985},
paper rupture
\cite {Kertesz/Horvath/Weber:1992},
and traveling salesman problem \cite{Kirkpatrick/Toulouse:1985}.
Though much effort has been devoted to study this problem, the general
solution is still lacking. There exist two approaches developed
recently to study this problem.
Cieplak {\it et al.\/}
\cite{Cieplak/Maritan/Swift/Bhattacharya/Stella/Banavar:1995} 
applied the max-flow algorithem for a two-dimentional energy landscape.
Another approach is to restrict the path to be directed,
that is, the path can not turn backwards. This approach is the directed
polymer problem which has been extensively studied in the past years,
see e.g.,
\cite{Perlsman/Schwartz:1992,Barabasi/Stanley:1995,Vicsek/Family:1991}.

In this manuscript, we adapt the {\it Dijkstra algorithm}
from graph theory
\cite{Cormen/Leiserson/Rivest:1990} for generating the optimal path
on a lattice with randomly distributed positive
energies assigned to the bonds. This algorithm enables us to generate
the optimal path between any two sites on the lattice, not restricted
to directed paths. We study the geometrical and
energetic properties of the optimal paths in $d=2$ and $3$ dimensions in random
uniform distribution of energies.
We calculate the scaling exponents for the width and the energy fluctuations  
of the optimal path. We find that for both
$d=2$ and $d=3$ the exponents are very close to those of directed polymers suggesting that the
non-directed optimal path (NDOP) is in the same universality
class as directed polymer (DP). Our results are in agreement with those 
found by Cieplak {\it et al.\/}
\cite{Cieplak/Maritan/Swift/Bhattacharya/Stella/Banavar:1995}
for the two dimentional case. This result indicates that in the case of uniformly
distributed energies the NDOPs are self affine and overhangs
do not play an important role in the geometry of NDOPs.

Our results are relevant, for example, in the following polymer realizations:
(i) Consider a d-dimension energy landscape in which there is a spherical regime of
randomly distributed high energies while outside this sphere the energies are zero or have very low
values.
Consider as well a polymer of length $N$ that one of its ends is 
attached to the center of the sphere while the other is free. The radius of 
the sphere is $r\ll N$ (see Fig 1a). The section of the polymer inside the sphere will reach the 
lowest energy path which is the optimal path studied here i.e., with a self
affine structure.
(ii) Consider a polymer in a d-dimensional energy landscape which is
divided into alternating strips of disordered low and high energies
(see Fig 1b).
In the strips of high energies the polymer is expected to behave like the 
optimal path.    

The Dijkstra algorithm enables one to find the optimal path from a
given source site to each site on a $d$-dimensional lattice.
During the execution of the algorithm, each site on the lattice
belongs to one of three sets (see Fig.~\ref{fig:fig2}):

\begin {itemize}
\item The first set includes sites for which their optimal path to the
source site have been already found.

\item The second set includes sites that are relaxed at least once but
their optimal path to the source has not been determined yet.
This set is the perimeter of the first set.

\item The third set includes all sites on the lattice which have
not been visited yet.
\end {itemize}

The algorithm itself consists of two parts, (i) initialization and
(ii) the main loop. The main loop, in its turn, is composed from
(a) the search and (b) the relaxation processes.

In the initialization part we prepare the lattice in the following way.
Each bond is assigned with a random energy value taken from a given
distribution.
Each site is assigned with an energy value of infinity.
We pick up a certain source site and assign it with the energy value
of zero and insert it into the second set.

After that we enter the main loop. We perform the search among
the sites from the second set and find the one with the minimal energy
value. Then we add it to the first set and proceed to the relaxation
process. This site is called the {\it added} site.
The relaxation process deals with sites neighboring to the
added site that do not belong to the first set.

In the relaxation process we compare two values, the energy value of the
neighboring site and the sum of two values: the energy value of the added site 
and the energy value of the bond between these sites.
If the value of the sum is smaller, then we
a) assign it to the neighboring site,
b) connect the neighboring and the added site by the path (thick bond
in the Fig.~\ref{fig:fig2}),
c) if the neighboring site belongs to the second set we break its previous
connection to another site (thick bond),
d) if the neighboring site does not belong to the second set we insert
it into the second set. 
The first four steps are demonstrated in Fig.~\ref{fig:fig2}.

Normally, the main loop stops when the second set is empty,
however one might wish to break the loop earlier, e.g., at the
moment when the first set reaches the edge of the lattice in
order to avoid boundary effects.

Each site that belongs to the first set is connected to the source by
a permanent path (thick bonds) that does not change during the execution
of the algorithm, so, if we stop the algorithm at any given time the
first set will still be valid.

We simulate both DPs and NDOPs on a square lattice in the following
way. Let $x, y$ be the horizontal and vertical axes. We choose the
origin to be the source site and study the optimal paths connecting
it with all the sites on the line between $[0,t]$ and $[t,0]$
for different values of $t$. The generalization to three dimensions is
straightforward. The random energies assigned to bonds are taken
from a uniform distribution. We find that our results are independent of
the distribution interval.

In Fig.~\ref{fig:fig3} we compare a configuration of DP and NDOP
on the same disordered energy landscape. 
It is seen that in the NDOP only very few overhangs exist. To test
the effect of the overhangs we calculate the mean end-to-end distance $R$
of the global optimal path (thick line in Fig.~\ref{fig:fig3})
as a function of its length $\ell$. The global optimal path is the
minimal energy path among all the paths with the same value of $t$.
Our numerical results clearly indicate the asymptotic relation $\ell \sim R$
showing that the NDOPs are self-affine \cite{Barabasi/Stanley:1995}.
We should compare this result to the strong disorder limit where
the paths can be regarded as self-similar fractals, with
$\ell \sim R^{d_{\rm opt}}$, where $d_{\rm opt} \simeq 1.22$ in $d=2$ and $d_{\rm opt} \simeq 1.42$ in $d=3$
\cite{Cieplak/Maritan/Banavar:1994/1996,Porto/Havlin/Schwarzer/Bunde:1997}.

In order to compare NDOP to DP we study several properties, such as
the roughness exponent $\xi$, the energy fluctuation exponent $\zeta$ for
two and three dimensions, as well as the distribution of the endpoints
of DP and NDOP. The above exponents are defined by the relations
$W \equiv \langle h^2 \rangle ^{1/2} \sim t^\xi$ and
$\Delta E \equiv \langle (E-\langle E\rangle)^2\rangle ^{1/2} \sim t^\zeta$.
Here, $h$ is the transverse fluctuation of the global optimal path
which is distance between its endpoint and the line $x=y$;
$E$ is the energy of the global optimal path which is the sum
of all bond energies along the path.
The average is taken over different realizations of randomness.
Fig.~\ref{fig:fig4} shows the dependence of width $W$ and energy
fluctuation $\Delta E$ of DP and NDOP on $t$ in two and three dimensions.
The points are the data for both DP and NDOP and the dashed lines
represent the exponents of the DP.
Our results indicate that the exponents for the NDOP are very close to those
of DP (see also Table~\ref{table:table1}).

Our results may be related to the recent findings \cite{Raisanen/Aeppala/Alava/Duxbury:1998}
that the roughness exponent of the minimal energy of domain wall in random Ising model and 
fracture interface are the same \cite{Kertesz/Horvath/Weber:1992} in 
$d=2$. In these cases, similar to our case, although overhangs may occur they do not play an important role.
  
In summary, our results suggest that the optimal path in the case of
uniformly distributed energies, for any energy interval, is in a different universality class
from the strong disorder limit but in the same universality class
as directed polymers. This result is relevant to several questions
regarding the equilibrium state of polymers in different realizations
of disordered energy landscape.

\widetext

\newpage

\begin {table}[p]
\caption{Width and energy fluctuation exponents of DP and NDOP in
$2$ and $3$ dimensions. The exponents were derived from the slopes
of the corresponding data points shown in 
Fig.~\protect\ref{fig:fig4}.
The error bar were estimated from taking five ensembles of
$10^4$ configurations each for $d=3$ and $500$ configurations
each for $d=3$.
}
\label{table:table1}
\begin {tabular}{|c|l|l|l|l|} \hline
        & \multicolumn{2}{c|}{$d=2$} & \multicolumn{2}{c|}{$d=3$} \\ \hline
        & \multicolumn{1}{c|}{DP} & \multicolumn{1}{c|}{NDOP} &
          \multicolumn{1}{c|}{DP} & \multicolumn{1}{c|}{NDOP} \\ \hline
\ $\xi$ \   & \ $0.66\pm 0.02$ \ & \ $0.67\pm 0.02$ \ &
\ $0.60\pm 0.05$ \ & \ $0.63\pm 0.05$ \ \\ \hline
\ $\zeta$ \ & \ $0.32\pm 0.02$ \ & \ $0.32\pm 0.02$ \ &
\ $0.19\pm 0.07$ \ & \ $0.18\pm 0.07$ \ \\ \hline
\end{tabular}

\end {table}

\begin{figure}[p]
\centerline{
\epsfysize=0.4\columnwidth{\epsfbox{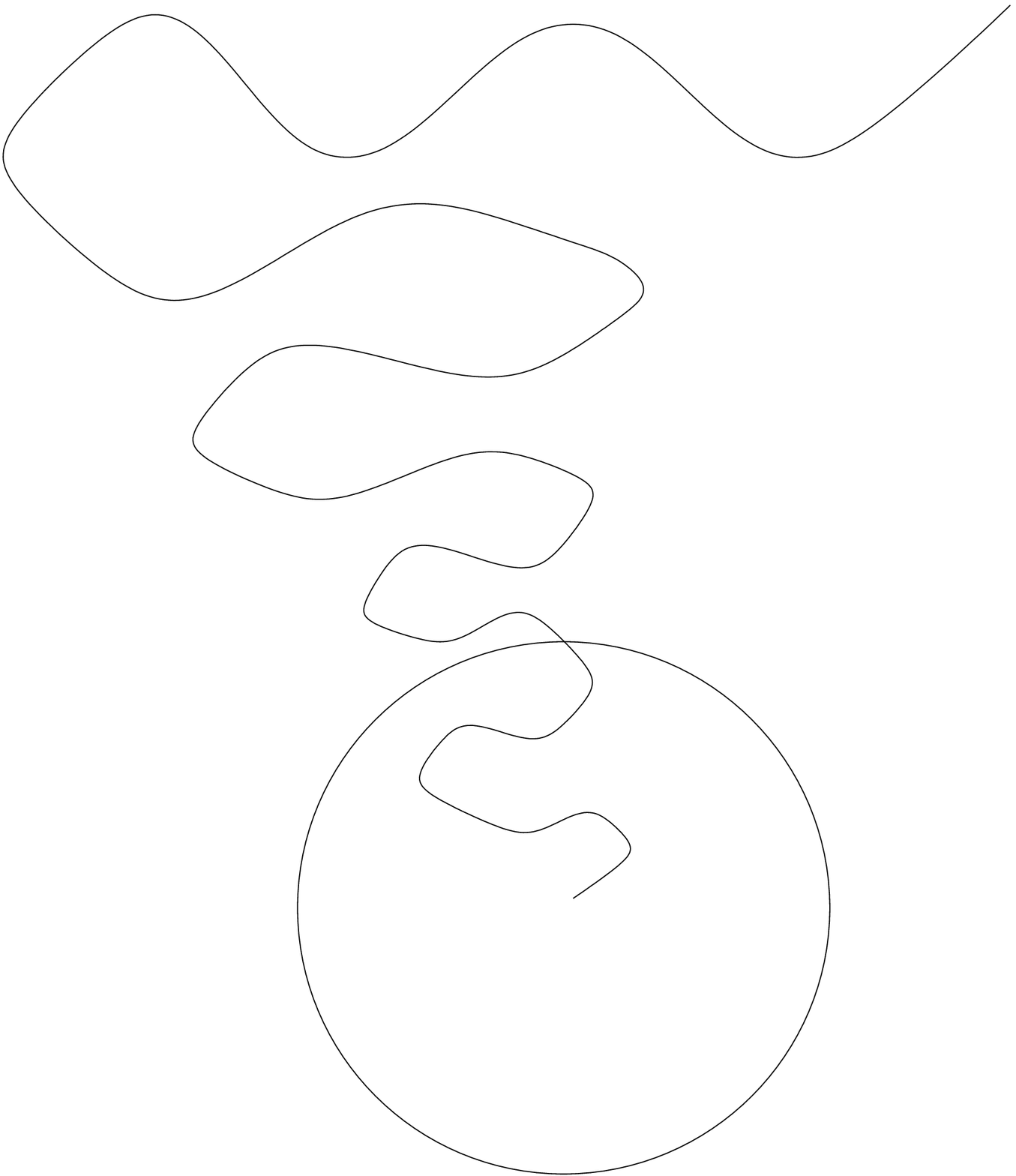}}
Fig.~1a
\hspace{1cm}
\epsfysize=0.4\columnwidth{\epsfbox{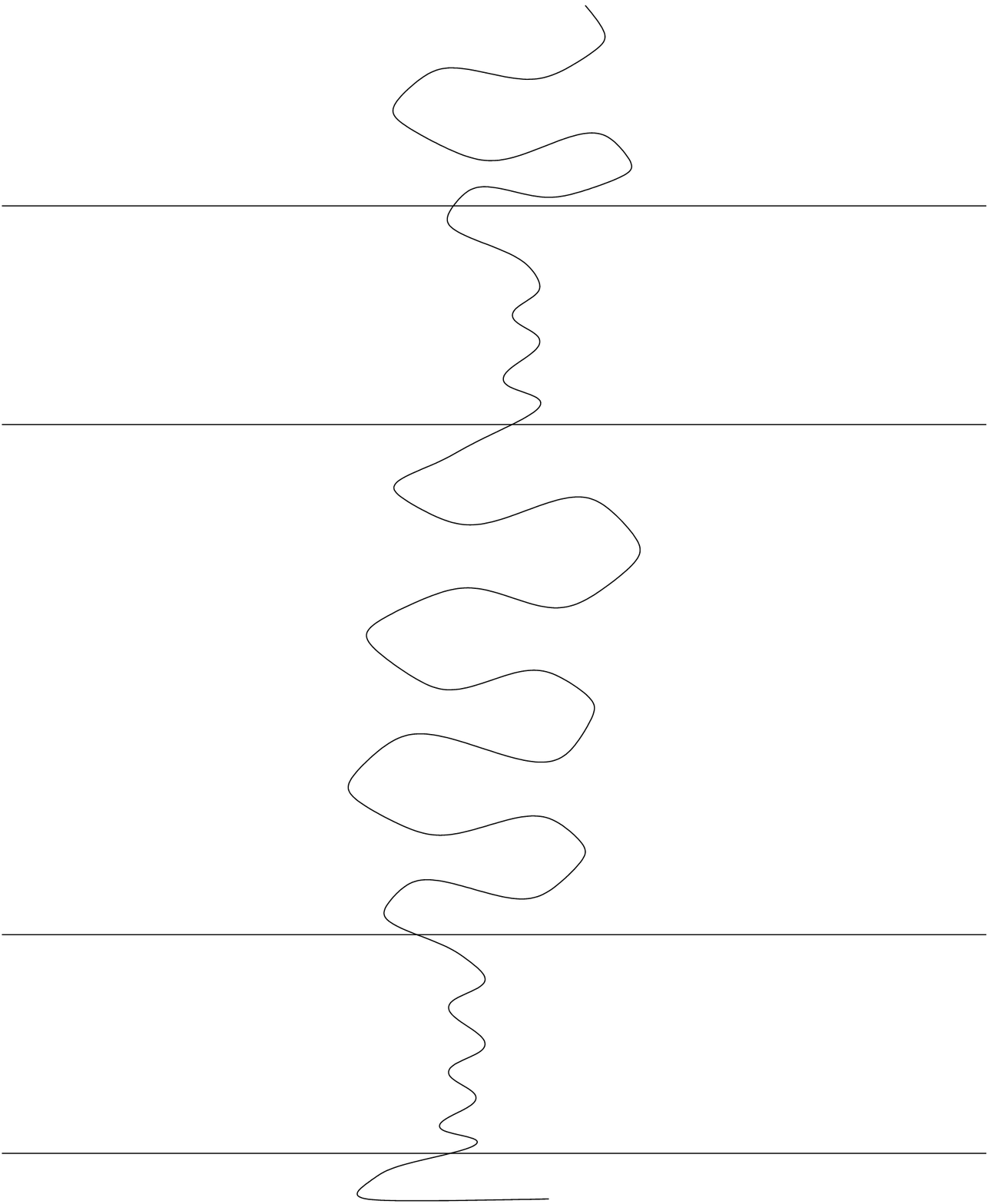}}
Fig.~1b
} 
\vspace{12pt}
\caption{Schematic illustrations of sections of a polymer having
the structure of optimal path.(a) within the circle energies are distributed
and outside the circle zero or very low energies are distributed.
(b) alternating strips of high and low distributed energies.
}
\label{fig:fig1}
\end{figure}

\begin{figure}[p]
\centerline{\epsfysize=0.5\columnwidth{\epsfbox{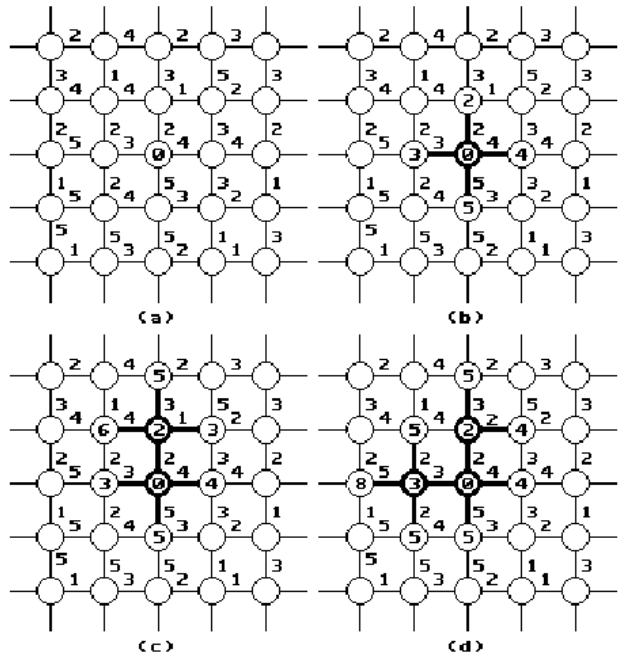}}}
\vspace{12pt}
\caption{Illustration of first four steps of the Dijkstra algorithm
applied to a square lattice. Numbers along the bonds represent the
random energy assigned to them. Numbers inside circles represent energies
of sites i.e., the total energy of the path connecting this site to
the source. Empty sites possess infinite energy and belong
to the third set, thick circles belong to the first set, all other
circles belong to the second set. Note e.g., that the site
marked $6$ in (c) got relaxed one more time and become $5$ in (d).
During this second relaxation we break its previous connection to the
site with energy $2$ and connect it to the site with energy $3$.
At each time step we identify the optimal path from each site in the first
set to the source by going along the thick bonds. 
}
\label{fig:fig2}
\end{figure}

\begin{figure}[p]
\centerline{\epsfysize=0.5\columnwidth{\epsfbox{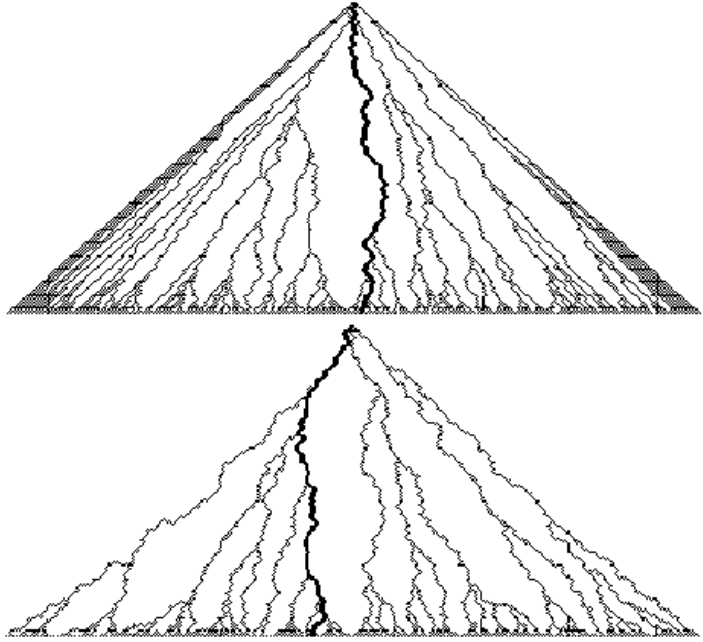}}}
\vspace{12pt}
\caption{The sets of all directed (the upper one) and non-directed
(the lower one) optimal paths with $t=300$ obtained for the same
realization of quenched randomness in the lattice. The global optimal
path which is the minimal energy path among all the paths with the
same $t$ is shown by a thick line. In this particular case the directed
and non-directed global optimal paths do not overlap. In other
cases they might overlap significantly though the rest of tree looks
somewhat different.
}
\label{fig:fig3}
\end{figure}

\begin{figure}[p]
\centerline{
\epsfysize=0.5\columnwidth{\rotate[r]{\epsfbox{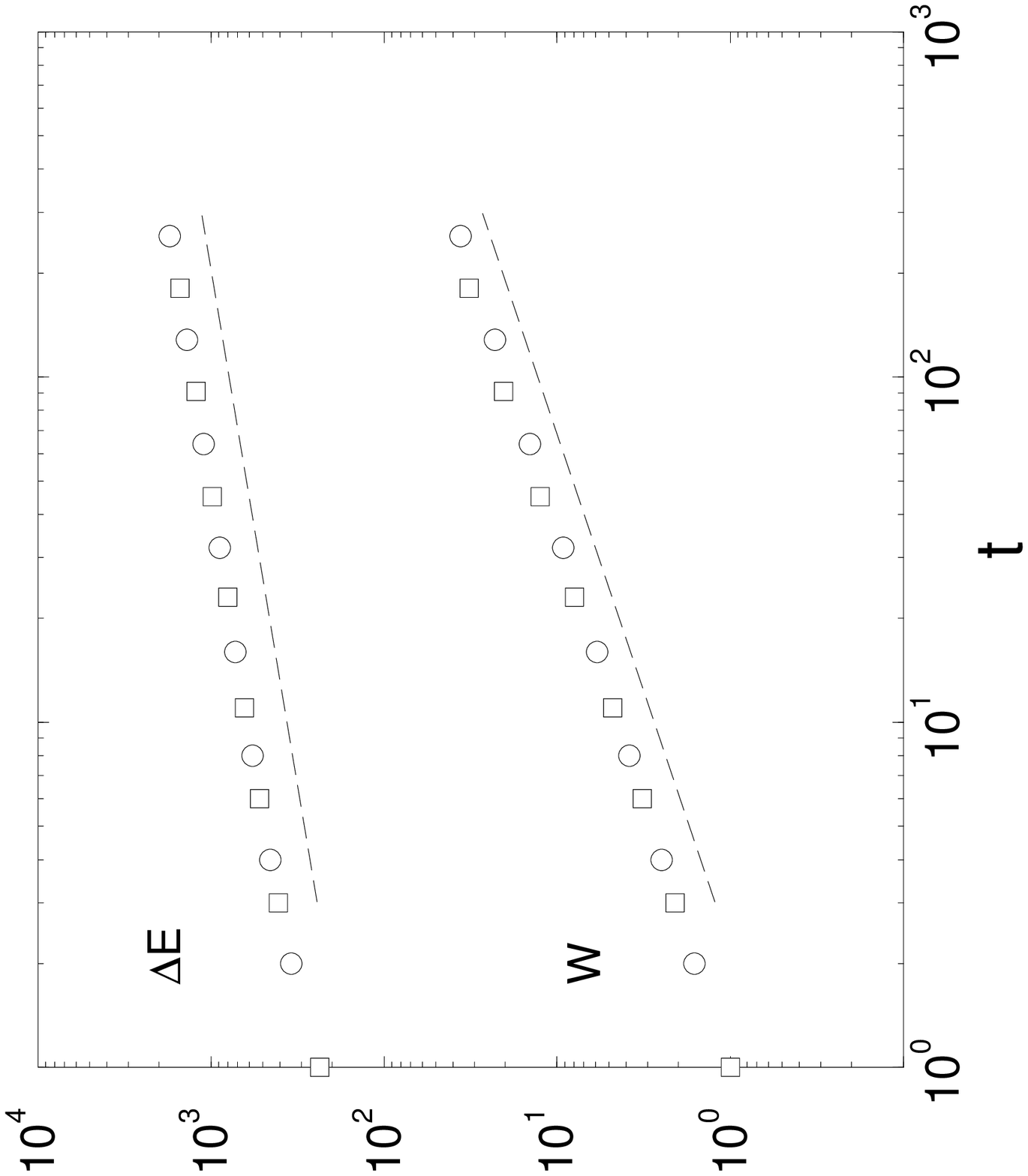}}}
Fig.~4a
\epsfysize=0.5\columnwidth{\rotate[r]{\epsfbox{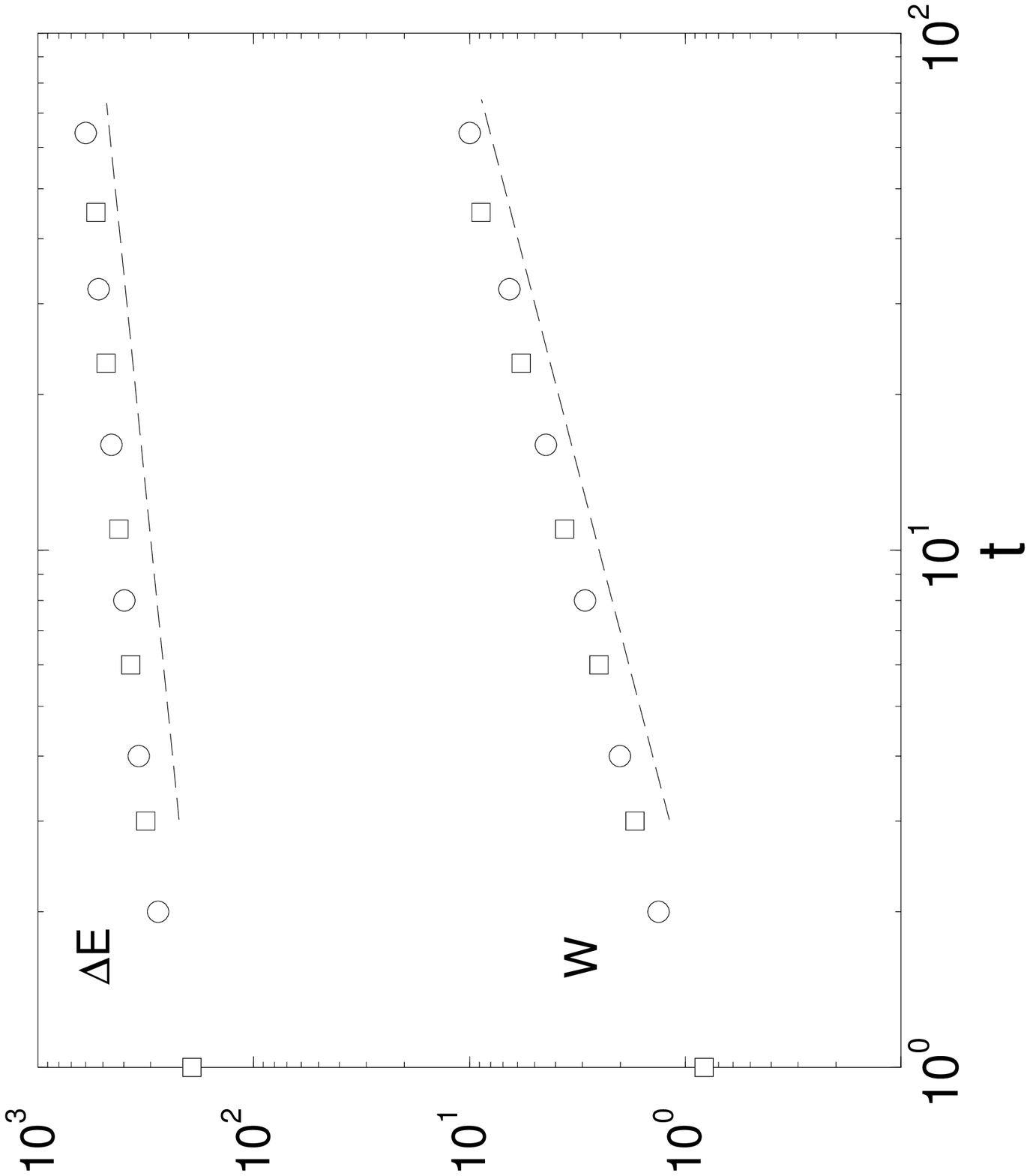}}}
Fig.~4b
} 
\vspace{12pt}
\caption{The width and the energy fluctuation as a function of $t$
on a double logarithmic plot in (a) two
and (b) three dimensions. Circles are used for directed polymers and
squares for non-directed optimal paths. For (a) $10^5$ systems
of linear size up to $t=300$ and for (b) $5000$ systems of linear
size up to $t=75$ are used. The dashed lines are given as a guide
to the eye and have slopes equal to the exponents known for DP:
$\xi=2/3$, $\zeta=1/3$ for $d=2$ and $\xi \simeq 0.62$,
$\zeta \simeq 0.24$ for $d=3$ \protect\cite{Perlsman/Schwartz:1992}.
For all cases  we used a uniform distribution of energies between
$E_1=1$ and $E_2=1000$. We also tested other energy intervals and found
the same results.
}
\label{fig:fig4}
\end{figure}

\newpage

\end{document}